\documentclass[bbm,amsfonts,amsmath,amssymb,12pt]{article} 
\usepackage[latin1]{inputenc} 
\usepackage[T1]{fontenc} 
\usepackage[english]{babel} 
\usepackage{amssymb,amsmath} 
\usepackage{amscd} 
\usepackage{fancyhdr} 
\usepackage{color} 
\usepackage{hyperref} 
 
\usepackage{epsfig} 
\usepackage{graphicx}
\usepackage{bm}
\usepackage{subfigure} 
 
\definecolor{red}{rgb}{1., 0., 0.} 
\definecolor{blue}{rgb}{0., 0., 1.} 
\definecolor{green}{rgb}{0.1, 0.7, 0.} 
\definecolor{purp}{rgb}{1,0,1}

\newcommand{\bc}{\begin{cases}\begin{aligned}} 
\newcommand{\ec}{\end{aligned}\end{cases}}

\newcommand{\de}{\partial}

\newcommand{\eq}{\begin{equation}} 
\newcommand{\fine}{\end{equation}} 
 
\newcommand{\dd}{\text d}

\def\drawbox#1#2{\hrule height#2pt 
        \hbox{\vrule width#2pt height#1pt \kern#1pt 
              \vrule width#2pt} 
              \hrule height#2pt}

\def\Asym#1#2{\vcenter{\vbox{\drawbox{#1}{#2} 
              \kern-#2pt       
              \drawbox{#1}{#2}}}}

\newcommand {\beq} {\begin{equation}} 
\newcommand {\eeq} {\end{equation}} 
 \newcommand{\be}{\begin{eqnarray}} 
\newcommand{\ee}{\end{eqnarray}} 
 
\newcommand{\KA}{K\"ahler} 
 
\def\re#1{(\ref{#1})} 
 
\begin{document} 
 
\begin{titlepage} 
 
\begin{flushright} 
\small 
\end{flushright} 
 
\vspace {1cm} 
 
 \begin{center}
\Large \bf  ${\cal N}=1$ Super Yang-Mills Domain Walls\\ via\\ The Extended Veneziano Yankielowicz Theory
 \end{center}
\vskip 1cm \centerline{\large P. Merlatti$^{a}$, F. Sannino$^{b}$, G. Vallone$^{c}$ 
and F. Vian$^{b}$} \vskip 0.1cm 
 
\vskip .5cm 
\begin{center} 
$^a$II. Institut f\"{u}r Theoretische Physik Universit\"{a}t Hamburg\\ Luruper Chaussee 149 D-22761 Hamburg, Germany \\~\\ 
$^b$The Niels Bohr and NORDITA, Blegdamsvej 17, DK-2100 Copenhagen \O, Denmark\\~\\  
$^c$ Dipartimento di Fisica Teorica, Universit\`{a} di Torino\\ and I.N.F.N., Sezione di Torino, via P.Giuria 1,
I-10125 Torino, Italy
\end{center} 
\vskip 1cm 
 
\begin{abstract} 
We investigate the vacuum structure of pure $SU(N)$ ${\cal N}=1$ super 
Yang-Mills. The theory is expected to possess $N$ vacua with 
associated domain walls. We show that the newly extended version of the low energy effective  
Lagrangian for super Yang-Mills supports the BPS domain wall solutions associated with any two vacua aligned with the origin of the moduli space. {}For the two color theory the domain wall analysis is complete. We also find new non BPS domain wall solutions connecting any two vacua of the underling SU(N) super Yang-Mills theory not necessarely aligned. When two vacua are aligned with the origin of the moduli space these solutions are the BPS ones. We also discuss the generic BPS domain wall solutions connecting any two vacua within the extended Veneziano-Yankielowicz theory. 
 
\end{abstract}

\end{titlepage} 
\section{Introduction} 
Understanding strong dynamics is a challenging and fascinating 
problem. Any progress in this area is likely to affect our 
understanding of Nature. Supersymmetric gauge theories are easier to 
understand than nonsupersymmetric ones and hence provide  
a natural 
laboratory for theoretical explorations. Actually, the higher is the 
supersymmetry involved, the more tractable the problem becomes. 
 
However, despite the enormous understanding gained on nonperturbative aspects of supersymmetric gauge dynamics \cite{Intriligator:1995au}, some puzzles remain unsolved. {}For example, since  pure ${\cal N}=1$ $SU(N)$ super Yang-Mills (SYM) is known to possess $N$ independent vacua, BPS domain wall solutions are expected to exist \cite{Dvali:1996xe}. Nonetheless, a proper domain wall description is still lacking. 

Disappointingly, even the low energy effective theory which was constructed long ago by Veneziano and Yankielowicz (VY) \cite{Veneziano:1982ah} does not support them \cite{Kovner:1997im,{Kovner:1997ca},{Chibisov:1997rc},{Smilga:1997cx}, 
{Witten:1997ep},{Kogan:1997dt}}\footnote{Another puzzling feature emerges when investigating these objects in the large $N$ limit of the underlying gauge theory. In this regime one may naturally expect their tension to scale as $N^2$, while it can be easily shown that it scales as $N$. This fact led to argue that they should be better viewed as D-brane like objects \cite{Witten:1997ep}.
 }. 

Perhaps alternative descriptions of ${\cal N}=1$ $SU(N)$ SYM via string theory or its low energy limit (i.e.  supergravity) may help shedding light on this subject. In recent years there has been much work in this area. {}For example, some nice results on the domain walls were obtained in \cite{Acharya:2001dz}, from the point of view of the three dimensional gauge theory living on them. We also recall that there are interesting descriptions of pure ${\cal N}=1$ $SU(N)$ SYM which are the gravitational duals found in 
\cite{Klebanov:2000hb,Maldacena:2000yy,Bertolini:2001gg} and further investigated in \cite{gg}. However, despite some qualitative as well as quantitative results found, not even within this framework a satisfactory description of the domain walls has been obtained.

In this paper we will investigate the domain walls issue within the effective Lagrangian approach, 
leaving the study of the string theory connection for a future publication.
A number of interesting models have been suggested in the literature \cite{Gabadadze:1998bi,deCarlos:1999xk} 
to address some of the problems related to the SYM domain wall solutions at the effective Lagrangian level.  
A common feature of all of these models is that new fields are needed to have a reasonable description of the domain wall solutions. 

In fact, the situation is even less satisfactory. Consider the case 
$N=2$; here not only the VY effective Lagrangian does not support a 
domain wall solution connecting the two vacua of the theory, but it 
actually displays domain wall solutions connecting the vacua with 
nonzero gluino condensate to the spurious vacuum 
\cite{Kovner:1997im,{Kovner:1997ca},{Smilga:1997cx}, 
{Kogan:1997dt}} 
with vanishing gluino condensate, {\it i.e.} the chirally symmetric 
one. This particular type of solution is expected to persist also at 
larger $N$ \cite{Kogan:1997dt}.

Here we investigate the issue of domain walls using a recently extended version of the VY theory\footnote{Some earlier attempts to generalize the VY theory to contain new degrees of freedom were considered in the literature using a three form supermultiplet\cite{Farrar:1997fn,{Cerdeno:2003us}}. The relation with the extended VY theory used here has been discussed in \cite{Merlatti:2004df}.} \cite{Merlatti:2004df}. We show that it is possible to find, at the effective Lagrangian level, domain wall solutions between the two vacua of $N=2$. The extended VY theory has a number of amusing properties investigated in detail in \cite{Merlatti:2004df,Feo:2004mr}. It contains a new chiral superfield with the same quantum numbers of scalar glueballs and its superpotential can be constrained using a number of consistency requirements.  
We will see that the introduction of this field is necessary for providing domain wall solutions between the two distinct vacua of the theory corresponding to nonvanishing gluino condensates.  
 
Another interesting point is that the requirement of the domain wall solutions being BPS will provide new insights on the otherwise completely unknown \KA~of the theory.  

We also find new non BPS domain wall solutions connecting any two vacua of SU(N) super Yang-Mills theory. Interestingly when two vacua are aligned with the origin of the moduli space these solutions are the BPS ones. We also discuss the generic BPS domain wall solutions connecting any two vacua within the extended Veneziano-Yankielowicz theory.

The paper is structured as follows: In the next section we will 
discuss the ${\cal N}=1$ domain walls in SYM. We then introduce the VY 
theory and summarize the associated domain wall problem discussed 
already in the literature. In section three we describe the extended 
VY effective theory and show how to recover the domain walls. We also introduce new solutions describing  
non-BPS domain walls associated to any two vacua of a generic $SU(N)$ SYM theory. In section four we analyze the issue of domain walls when two fields are present. We also discuss the problem of the BPS domain wall solutions connecting vacua not aligned with the origin in section five and 
show that there is a possible intriguing relation with the domain wall solutions for supersymmetric QCD. 
Conclusions are presented in the last section. Two appendices are provided. In appendix A, for reader's convenience, we summarize the BPS conditions for domain walls for a generic nonlinear supersymmetric sigma model. We investigate, in appendix B, the effects on the domain wall solutions when a different \KA~ is used.   
 
\section{Brief Review of the Domain Walls in SYM and the VY Theory} 
 
In $\mathcal N=1$ SYM there are  $N$ distinct vacua labelled by the value of the gluino 
condensate which, in our normalization, reads $\langle\mbox{Tr}\lambda\lambda\rangle \propto N 
\Lambda^3 
\exp (2\pi ik/N)$.  
Topologically stable domain walls interpolating between different vacua are expected. Their tension \cite{Dvali:1996xe} should be 
\begin{equation} 
\label{gauginocond} 
\mathcal E= \frac 
N{8\pi^2}\vert 
\langle\mbox{Tr}\lambda^2\rangle_{\infty}-\langle\mbox{Tr}\lambda^2\rangle_{-\infty}\vert \,, 
\end{equation} 
where the subscript $\pm \infty$ marks the values of the gluino 
condensate at $z\to\pm\infty$. These domain walls are BPS saturated 
configurations which preserve 1/2 of the original supersymmetry \cite{Dvali:1996xe}.  
 
Explicitly, the domain wall tension has the following $N$ dependence 
\begin{eqnarray} 
N^2\sin\frac{\pi k}{N} \ ,  
\end{eqnarray} 
where $2\pi k/N$ is the phase difference between two generic vacua.  
 
When $k$ is fixed and $N$ becomes large the tension is O($N$). However, when $k$ scales with $N$  the tension becomes O($N^2$) and the associated domain walls should be described as 
soliton--like solutions in an effective theory  \cite{Witten:1997ep}. On the other hand, Witten argued that a more suitable interpretation of the domain walls whose tension scales like $N$ could be  
in terms of D-branes-like objects \cite{Witten:1997ep}.  
 
We now review the present status of the domain wall solution for pure SYM
via the VY theory, which concisely summarizes the symmetry of the 
underlying theory in terms of the composite chiral superfield $S$ 
\begin{eqnarray} 
S=\frac{3}{32\pi^2\, N}{\rm Tr}W^2 \ , 
\end{eqnarray} 
where $W_{\alpha}$ is the supersymmetric field strength. When 
interpreting $S$ as an elementary field it describes a gluinoball 
and its associated fermionic partner. 
 
The K\"ahler potential and the superpotential are respectively \cite{Veneziano:1982ah} 
\eq\label{VY superpotential} 
\mathcal K(S,\bar S)=\frac{9N^2}\alpha(S\bar S)^{\frac13}\,,\qquad 
W_{VY}[S]=\frac{2N}{3}S\log_{(0)} 
\left(\frac{S}{e\Lambda^3}\right)^N\,, 
\fine 
where $\alpha$ is a dimensionless real parameter and we restricted ourselves to the first branch of the logarithm. One determines the  
$N$ vacua $S=\Lambda^3e^{\frac{2k\pi i}{N}}$  of the theory labelled by $k=0,\ldots N-1$ by extremizing the superpotential, namely 
 \begin{eqnarray} 
 \frac{\de W_{VY}}{\de S}=0 \ .  
 \end{eqnarray}  
To investigate the domain wall solutions it is sufficient to study the bosonic Lagrangian of the theory 
\eq \label{bosoniclagrangian} 
L_{VY}=-\frac{N^2}\alpha(\varphi\bar\varphi)^{-\frac23}\de_\mu\varphi\de^\mu\bar\varphi+\frac{4}{9} 
\alpha(\varphi\bar\varphi)^{\frac{2}{3}}\left|\log_{(0)}\left({\varphi}\right)^N\right|^2 \,, 
\fine 
having written $S=\varphi +\sqrt{2} \theta \psi  + \theta^2F$ and from now on we will set $\Lambda=1$ to ease the notation. 
We note that although the superpotential \eqref{VY superpotential} 
is discontinuous on the rays $S \propto e^\frac{(2k+1)\pi i}{N}$, 
the potential $\left|\log_{(0)}\left(\frac{\varphi}{\Lambda^3}\right)^N\right|^2$ 
is continuous at $\varphi \propto e^\frac{(2k+1)\pi i}{N}$ (while its derivative is not). 
 
It was noticed in \cite{Kovner:1997im} that the effective theory has a 
vanishing potential not only for the vacua 
$\varphi=e^{\frac{2k\pi i}{N}}$ but also for the value 
$\varphi=0$. There are, however, theoretical arguments which do not 
support the existence of such a vacuum state 
\cite{Cachazo:2003yc}. Besides, one should notice that since the 
kinetic term is divergent for $\varphi\rightarrow 0$, one is not really 
allowed to consider such a vacuum\footnote{One may argue that by 
renaming the field $\varphi$ with $\varphi^3$ the kinetic term becomes 
a canonically normalized term while the potential still has a zero at 
the origin. In this case we expect, due to Haag's theorem, the field 
redefinition to provide the same physics in a given vacuum. However 
the global structure of the minima changes dramatically with the field 
redefinition and special attention must be paid. This is true 
especially at the origin of the field space since with the original 
field definition the kinetic term is singular.}. Setting aside the problems associated with the extra vacuum at the origin of the moduli space, one can still look for domain wall solutions in this theory. {}As explained in  appendix A, we can write for the case at hand the following linear differential equation 
\eq\label{BPS VY} 
\frac{\dd\bar\varphi}{\dd z}=  
e^{i\beta}  \frac{2\alpha}{3N}  
(\varphi\bar\varphi)^{\frac{2}{3}}\log_{(0)}\varphi^N\;, 
\fine 
where  $\beta$ is an arbitrary phase. 
 
If we consider the case $N=2$, it is then possible to find real solutions for $\varphi$ interpolating between one of the vacua at $\varphi=\pm 1$ and the spurious one at $\varphi=0$ \cite{Kovner:1997ca}. This is clearly a strange situation in which we expect the origin not to be a true vacuum while the BPS domain wall solutions are well-defined only between the true vacua and the origin and no solution exists interpolating between $\varphi=\pm 1$ \cite{{Kovner:1997ca},{Smilga:1997cx}, 
{Kogan:1997dt},{Smilga:1997yp}}. We now show that such solutions can be found when using the extended VY theory \cite{Merlatti:2004df}.  

\section{Extended VY Theory versus Domain Walls} 
In the previous section we reviewed the failure of the simple VY effective Lagrangian to  
describe the expected domain solutions. To overcome these difficulties, also guided by previous insights \cite{Kogan:1997dt}, it is natural to think that new degrees of freedom are needed. {}Indeed, even before considering the domain wall solutions, it was shown that the introduction of new states with zero $R$ charge in the VY theory is important. 
{}For example these states are relevant when breaking supersymmetry by 
adding a gluino mass term \cite{Merlatti:2004df}. This is so since 
the basic degrees of freedom of the pure Yang-Mills theory are 
glueballs. {}Further support for the relevance of such glueball 
states in SYM
comes from lattice simulations 
\cite{Feo:2002yi}. This state is encoded in a new chiral superfield $\chi$ with the proper quantum numbers. Moreover it was possible to motivate a specific form for the extended VY superpotential 
which has a number of amusing properties. The introduction of a new field  has also a natural counterpart in the 
geometric approach to the effective Lagrangian theory proposed by 
Dijkgraaf and Vafa \cite{Dijkgraaf:2002dh}.  
 
In \cite{Merlatti:2004df} the focus was on the properties of the 
theory depending solely on the superpotential, which reads 
\begin{eqnarray}\label{ms} 
W\left[S,\chi\right] &=& 
\frac{2N^2}{3}S\left[\log_{(0)}S -1 -\log 
\left(-e\frac{\chi}{N} \log_{(0)}\chi^N\right) \right] \ , 
\end{eqnarray} 
where $S$ is the gaugino bilinear superfield defined after (\ref{bosoniclagrangian}) 
and $\chi$ describes the $R=0$ 
glueball type degrees of freedom ($\chi=\varphi_{\chi} + 
\sqrt{2}\theta \psi_{\chi}  + \theta^2 F_{\chi}$). 
 
\begin{figure}[htbp] 
\begin{center} 
\includegraphics[scale=.6]{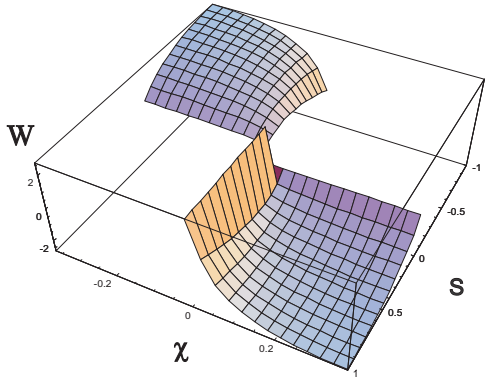} 
\end{center} 
\caption{Superpotential of the extended VY theory for the case $N=2$.} 
\label{butterfly} 
\end{figure}

For illustration, in figure \ref{butterfly} we plot the 
superpotential of the theory in the $N=2$ case.  
There are two extrema (the points where the first derivative of the superpotential vanishes), {\it i.e.}  $(S=1, \chi=1/e)$ and $(S=-1, \chi=-1/e)$. Note that before taking into account the \KA~we have no extremum at the origin of the moduli. Here we do not distinguish, labelwise, between the chiral superfields and their bosonic components.  
 
A K\"{a}hler potential is, in general, needed to investigate dynamical properties as done in \cite{Feo:2004mr}. VY suggested the simplest 
K\"{a}hler for $S$ which does not upset 
the saturation of the quantum anomalies,  {\it  i.e.} $( S\bar{S})^{1/3}$. However, due to the 
presence of the new field one can modify the VY K\"{a}hler in order 
to provide a kinetic term also for $\chi$ as follows 
\begin{eqnarray} 
\label{kahler} 
K(S,\bar{S},\chi,\bar{\chi})~=~ 
\frac{9\,N^2}{\alpha}(S\bar{S})^{1/3}h(\chi,\bar{\chi})\ , 
\end{eqnarray} where $h(\chi,\bar{\chi})$  
is a generic real positive definite function of $\chi$  and $\bar\chi$. We stress that since $\chi$ is a dimensionless field and in order not to upset 
the saturation of the quantum anomalies, at the effective Lagrangian level, the generic form of 
the new K\"{a}hler for $S$ and $\chi$ is the one presented above. The associated K\"{a}hler metric is 
\begin{eqnarray} 
\label{matrix} 
g^{i\bar{\jmath}} = \left[g_{i\bar{\jmath}}^{-1}\right]^T \ , 
\quad {\rm with} \quad g_{i\bar{\jmath}}=\frac{\partial^2\,K}{\partial 
\varphi^i\partial \bar{\varphi}^{\bar{\jmath}}} \ , \end{eqnarray} and 
$\varphi^1=\varphi$ while  
$\varphi^2=\varphi_{\chi}$.  
The potential is then 
\begin{eqnarray}V\left[\varphi\ ,\varphi_{\chi}\right]= \frac{\partial W}{\partial \varphi^i}\, 
g^{i\bar{\jmath}}\, \frac{\partial {\overline W}}{\partial 
\bar{\varphi}^{\bar{\jmath}}}\ .\end{eqnarray} 
 
{}Note also that since eliminating $\chi$ via its equation of motion at the 
superpotential level 
\begin{eqnarray} 
\frac{\partial W\left[S,\chi\right]}{\partial \chi} = 0 \, , 
\end{eqnarray} 
yields 
\begin{eqnarray} 
\chi = {e^{\frac{2i\,\pi k}{N} }}\,e^{-1} \ , 
\end{eqnarray} 
we both recover the effective Lagrangian of the VY with an opportunely 
redefined $\alpha$ and in a very natural way the corrections advocated by Kovner and 
Shifman \cite{Kovner:1997im} emerge. It is now natural to try to use the extended VY theory  
to overcome some of the problems related to the domain wall solutions of the VY theory.  
 
\subsection{Domain Walls in the Extended VY} 
Before showing the complete solution for two fields, it is instructive 
to display the domain wall solution for $\chi$ after having 
integrated out $S$ via its equation of motion. This will be helpful in 
building up our intuition on the more general and complicated 
solution featuring the two fields. If any domain wall solution is found we simply reconstruct 
the domain wall for $S$ via its functional dependence on $\chi$. We 
will see that, while there are relevant corrections when solving for the 
complete two field theory, the basic picture remains 
unchanged. When integrating out $\chi$ via its equation of 
motion we return to the VY theory improved {\it a l\'a} Kovner and Shifman \cite{Kovner:1997im}, for which no solution interpolating between the desired vacua is found. 
 
We will start from a general \KA~for $\chi$ and require a functional 
form which is compatible with the presence of BPS saturated domain wall 
solutions. It is really the freedom in choosing the \KA~for $\chi$ 
which allows us to find the desired domain wall solutions.  
There is no analog of this in the case of the simple VY theory, where instead the \KA~is constrained for dimensional reasons.  
 
The equation of motion for $S$ reads 
\begin{eqnarray} 
\frac{\partial W\left[S,\chi\right]}{\partial S} = 0\,, 
\end{eqnarray} 
yielding 
\begin{eqnarray} 
\label{outs} 
S&=&-\frac{e}{N}\,\chi\, 
\log_{(0)} 
\chi^N \,  \ , 
\end{eqnarray} 
which at the bosonic level can be used to relate the domain wall solution for $\varphi_{\chi}$ to $\varphi$. 
The effective theory for $\chi$ is then 
\begin{eqnarray} 
\label{eff1} 
{\cal L} &=& \int d^2 \theta 
d^2\bar{\theta}\, g(\chi,\bar{\chi})   +\frac{2N}{3}e\,\int\, 
d^2\theta\, \chi \log_{(0)}\chi^N + {\rm c.c.} \,, 
\end{eqnarray} 
where $g(\chi,\bar{\chi})$ is obtained from \re{kahler} via the substitution \re{outs} 
\begin{eqnarray} 
g(\chi,\bar{\chi}) = \frac{9 N^2}{\alpha} (S(\chi)\bar{S}(\bar{\chi}))^{\frac{1}{3}}h(\chi,\bar{\chi}) \ . 
\end{eqnarray} Integrating out $S$ in this way, strictly speaking, is valid only in the limit in which $S$ is much heavier than $\chi$. Recently it has been demonstrated that this is not the case \cite{Feo:2004mr}, so this section must be considered only as a simple warm-up exercise which will help us understand the more general result we will obtain in the following section.   
 
\begin{figure}[htbp] 
\begin{center} 
\includegraphics[scale=.6]{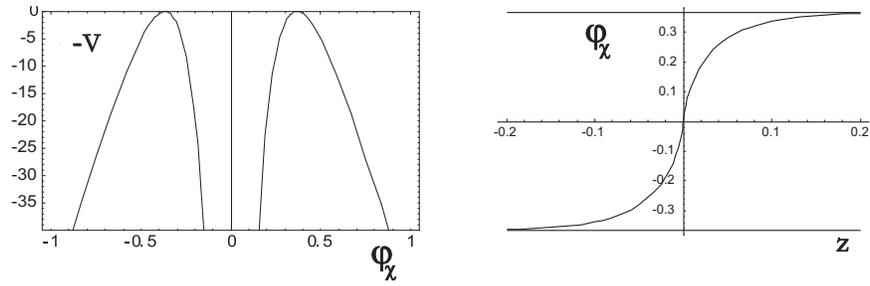} 
\end{center} 
\caption{Potential of the theory with a simple \KA~for $\chi$ and the 
associated explicit domain wall solution obtained by setting $\rho =1$ in the $N=2$ case. } 
\label{schi} 
\end{figure}

\begin{figure}[htbp] 
\label{simples} 
\begin{center} 
\includegraphics[scale=.6]{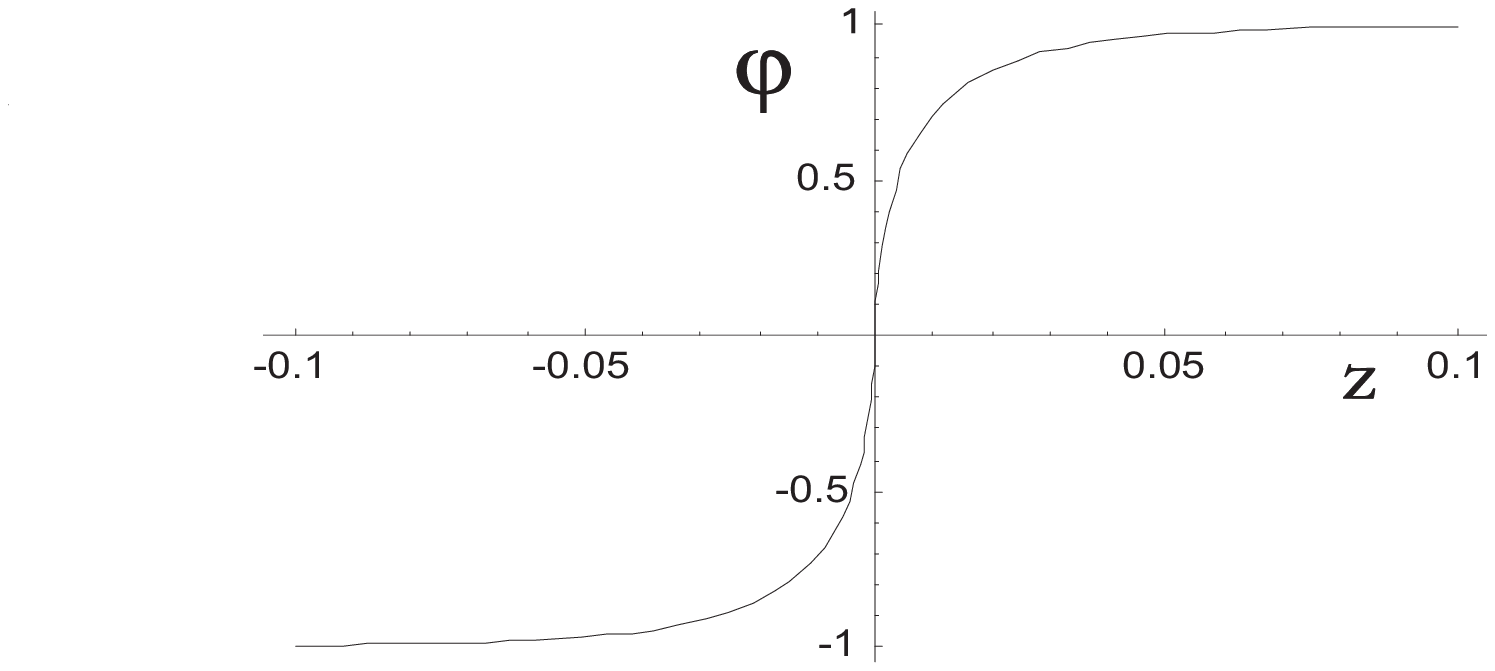} 
\end{center}
\caption{Domain wall solution for $S$ obtained using the equation of motion linking $S$ and $\chi$. } 
\end{figure}

Having derived the Lagrangian for the effective theory, we can now write the BPS domain wall differential equation specializing \re{BPS solution} in the appendix A to the case of a single field 
\begin{eqnarray} 
\label{BPS1}  
\frac{\dd\varphi_{\chi} }{\dd z}= e^{-i\beta}  
g^{\chi \bar{\chi} } 
\de_{\bar{\chi}}{\overline W} \,, 
\end{eqnarray} 
where the metric can be read from \re{matrix} and it is understood 
that derivatives and \KA~metric indices are evaluated on the bosonic 
sector of the theory. Here $\beta$ is a generic phase.    
{}Using equation (\ref{eff1}) we know that:
\begin{eqnarray} 
 \partial_{\bar{\chi} }{\overline W}= 
\frac{2N}{3} \,e  
\left(\log_{(0)} \bar{\varphi}_\chi^N + N\right)\,. 
\end{eqnarray} 
Problems arise since $S\propto \chi\log\chi$ and, in general, the \KA~ leads to  
very singular kinetic terms at the origin of the moduli space. These problems can be 
circumvented if one suitably chooses  the function $h(\chi,\bar{\chi})$ 
so as to  satisfy a number of constraints which will be explained in 
much more detail in the next section. For illustration it is 
sufficient to device $h$ such that one has the canonical \KA~for 
$\chi$, {\it  i.e.} $\chi\chi^{\dagger}/\rho$, with $\rho$ a real 
parameter. 
We then have the following first order differential equation: 
\begin{eqnarray} 
\frac{\dd\varphi_{\chi}}{\dd z} = e^{-i\beta} \rho \frac{2N}{3} \,e  
\left(\log_{(0)} \bar{\varphi}_\chi^N + N\right)\,. 
\end{eqnarray} 
This amounts to two independent differential equations, one for the 
modulus and the other for the phase of $\varphi_{\chi}$. After defining 
\begin{eqnarray}\varphi_{\chi}= \eta \,e^{i \delta} \ ,\label{phicdef}
\end{eqnarray} they read 
\begin{eqnarray} 
\label{mod}  
\frac{d\eta }{dz} &=& \rho\frac{2N}{3}e\left[\cos(\beta + \delta)(\log\eta^N +N) - \sin(\beta + \delta) (N\delta)_{(0)}\right] \\ 
\label{phase}  
\eta\frac{d \delta }{dz} &=& \rho\frac{2N}{3}e\left[-\sin(\beta + \delta)(\log\eta^N +N) - \cos(\beta + \delta) (N\delta)_{(0)}\right] \,, 
\end{eqnarray} 
where 
\eq 
(N\delta)_{(0)}=\delta N-2 k\pi\ , \qquad\text{if}\quad -\pi+2k\pi<\delta N< \pi+2k\pi \ . 
\fine 
Consider the case $N=2$. The domain wall solution we would like to 
find must connect the two vacua at $\chi=\pm 1/e$. We can look for a 
solution passing through the origin of the moduli space. In this case 
we have that the phase of $\varphi_{\chi}$ is $\pi$ for $z\in \,]-\infty,0]$ and is $0$ for 
$z\in \,]0,\infty [$. We split the equations in two parts, according to the range of the coordinate $z$. {}For $z\in \, ]-\infty, 0]$ we have: 
\begin{eqnarray} 
\frac{d\eta }{dz} &=& \rho\frac{4}{3}e(\log\eta^2 +2) \ , 
\end{eqnarray}   
 having set $\beta=\pi$, whereas 
\begin{eqnarray} 
\frac{d\eta }{dz} &=& -\rho\frac{4}{3}e(\log\eta^2 +2) \ , 
\end{eqnarray} for $z\in \, ] 0, \infty[$. It is easy to solve this 
differential equation. Clearly, in the former the modulus decreases 
from $1/e$ to zero while in the latter it increases from $0$ to $1/e$. In figure \ref{schi} we portray the potential for the case $N=2$ and the solution for the classical configuration.  
Since the superpotential is continuous at the origin and we used a simple \KA, the standard BPS formula (\ref{tension}) for the tension applies yielding:  
\begin{eqnarray} 
\mathcal E = 2|W_{+\infty} - W_{-\infty}|=\frac{32}{3} . 
\end{eqnarray} 
Here $W_{\pm\infty}$ is the superpotential of the theory 
evaluated at the two different minima reached at $z=\pm\infty$. We 
note that the domain wall tension remains unaltered even after having 
integrated out $S$. This is another test of the extended 
VY \cite{Merlatti:2004df}, {\it  i.e.} not only we can eliminate the 
troublesome vacuum at the origin of the moduli space, but we can now 
have a domain wall solution interpolating between the two vacua with 
the expected tension. 
Finally, the domain wall for $S$ can be 
easily reconstructed via the equation of motion for $S$ as function of $\chi$, as shown in figure \ref{simples}.
\subsection{Non-BPS Domain Walls} 
In the underlying theory the domain wall tension between two vacua differing by $k$ 
reads 
\begin{eqnarray} 
\label{sine} 
\mathcal E = \frac{8}{3}\, N^2\,\sin\pi\frac{k}{N} \ . 
\end{eqnarray} 
Consider two cases: 1) $k$ is fixed and $N$ becomes large. Here the tension is O($N$). 2) $k$ scales with $N$ and hence the tension is  
O($N^2$).  
{}For an even number of colors we can find 
immediately domain wall solutions between two vacua lying on a straight line 
crossing the origin, {\it i.e.} $k=N/2$. These simple generalizations of the $N=2$ 
case are found by wisely choosing $\beta$, and have tension of order $N^2$.  
We recall that domain wall solutions whose tension is of order $N^2$ are naturally expected to emerge as solitonic solutions of a low energy effective theory description \cite{Witten:1997ep}. However our solutions do not exhaust all of the possible domain walls with tension of order $N^2$.

The situation is more involved when two vacua are not aligned with the 
origin. {}For the VY theory it has already been pointed out in \cite{Kogan:1997dt} that one has 
to be very careful in trying to find solutions which do not pass 
through the origin when the effective lagrangian has a structure with 
$N$ distinct sectors in the space of fields. The superpotential happens 
to be singular along the boundaries of a sector while the potential of the theory is obtained by gluing together 
different potentials along the boundaries, and cusps develop.  
 
Here we face a similar problem since the extended VY theory also 
contains cusps, but now in the variable $\chi$. According to \cite{Kogan:1997dt}, by resolving the  problem of cusps, one might hope to obtain the expected O($N$) tension between adjacent vacua. The claim is supported by a number of toy models and the final solution might even require the introduction of $N$ fields. We will discuss in the last section these kind of domain walls within the extended VY theory.

We now investigate domain wall solutions interpolating between two generic vacua but still  
passing through the origin. This can be achieved by simply choosing two different $\beta$'s, 
one for each line connecting one of the two vacua to the origin. Since there are two different $\beta$'s, solutions will not preserve, in general, ${\cal N}=1/2$ supersymmetry and these objects will not be BPS saturated. 
 
{}We start by considering two vacua differing by a phase of $2\pi k/N$ in the gluino condensate. {}For $z\in \,  
]-\infty, 0]$, using the definition in eq.~(\ref{phicdef}), we fix $\delta=2k\pi/N$, while $\delta=0$ for $z\in \, ]0, \infty[$. According to  
(\ref{mod}), (\ref{phase}), in the range  $z\in \, ]-\infty, 0]$, we have the following differential equations: 
\begin{eqnarray} 
\frac{d\eta }{dz} &=& \rho\frac{2N}{3}e\left[\cos(\beta_1 + \frac{2k\pi}{N})(\log\eta^N +N) \right] \ ,\\ 
0 &=& \rho\frac{2N}{3}e\left[\sin(\beta_1 +\frac{2k\pi}{N})(\log\eta^N +N)\right] \ . 
\end{eqnarray} 
The second equation demands $\beta_1 = -2k\pi/N + n\pi$, while the condition of a decreasing modulus fixes $n$ to be even. In the region $z\in ]0,\infty[$ we have $\delta=0$  
and hence 
\begin{eqnarray} 
\frac{d\eta }{dz} &=& \rho\frac{2N}{3}e\left[\cos\beta(\log\eta^N +N) \right] \ ,\\ 
0 &=& \rho\frac{2N}{3}e\left[\sin\beta(\log\eta^N +N)\right] \ . 
\end{eqnarray} 
The second equation requires $\beta=n\pi$ and the first that $n$ is odd to ensure that the modulus increases with $z$. These differential equations are identical to the one solved for the $N=2$ case. To reconstruct the complete solution one simply glues the two solutions at the origin of the moduli space. Due to the fact that $\beta$ is constant in the two different regions the domain wall tension is  
\begin{eqnarray} 
2\left({\Re}e[e^{i\beta}W_{\infty}] - {\Re}e[e^{i\beta}W[0]]+ {\Re}e[e^{i\beta_1}W[0]]- {\Re}e[e^{i\beta_1}W_{-\infty}]\right)=8\frac{N^2}{3} \ . 
\end{eqnarray} 
This must be confronted with the sine formula (\ref{sine}) which shows that, using the underlying theory, the BPS and ${\cal N}=1/2$ associated solution has a lower tension. The bound is saturated by the domain wall solution with $N$ even and 
$k=N/2$. These are exactly the solutions we  found earlier. 
 
Our construction cannot be reproduced using the simple VY 
theory. This is so since in the VY theory there are no domain wall 
solutions connecting two chirally asymmetric vacua and passing through 
the origin. 
 
 
 
The present analysis has been performed in the simple case in which we integrated out $S$. Since we will show that it is possible to have a  
solution when keeping $S$, the results are valid in general.  
 
\section{The Two Field Problem} 
We now consider the two field problem. The situation is involved since the fields not only interact at the superpotential level but also due to the K\"ahler. Here we will derive some general {\it constraints} on the  K\"ahler structure of the effective theory by requiring the domain wall solutions to exist and to be BPS saturated.

We first recall the BPS domain wall differential equations for a generic supersymmetric nonlinear sigma model with $D$ fields, presented in the appendix (see \re{BPS solution}) and reported here for the reader's convenience 
\eq 
\frac{\dd\varphi^i}{\dd z}=e^{-i\beta} g^{i\bar\jmath}\de_{\bar\jmath}{\overline 
W} \,,\qquad i=1,\ldots,D \,.\fine  
We have only the two fields $\varphi^1=\varphi$ and $\varphi^2=\varphi_{\chi}$ and the system of differential equations reduces to 
\begin{eqnarray} 
\frac{\partial}{\partial z}\left(%
\begin{array}{c} 
   \varphi\\ 
  \varphi_{\chi} \\ 
\end{array}%
\right) = \frac{\alpha}{N^2} \frac{(\bar{\varphi}\varphi)^{\frac{2}{3}}}{h\, h_{\chi\bar{\chi}} - h_{\chi}\bar{h}_{\bar{\chi}}} \left[%
\begin{array}{cc} 
  h_{\chi\bar{\chi}} & -\frac{{\bar{\varphi}^{-1}}h_{\chi}}{3} \\ 
  -\frac{\varphi^{-1}\bar{h}_{\bar{\chi}}}{3} & \frac{h}{9}  ({\bar{\varphi}\varphi})^{-1}\\ 
\end{array}%
\right] \,\left(%
\begin{array}{c} 
  \partial_{\bar{S}}{\overline{W}} \\ 
   \partial_{\bar{\chi}}{\overline{W}}\\ 
\end{array}%
\right)\,, \label{DE} 
\end{eqnarray} 
with $h_{\chi} = \partial_{\chi}h$, $\bar{h}_{\bar{\chi}}=\partial_{\bar{\chi}}h$ and $h_{\chi\bar{\chi}}= \partial_{\chi}\partial_{\bar{\chi}}h$.

\begin{figure}[htbp] 
\begin{center} 
\includegraphics[scale=.6]{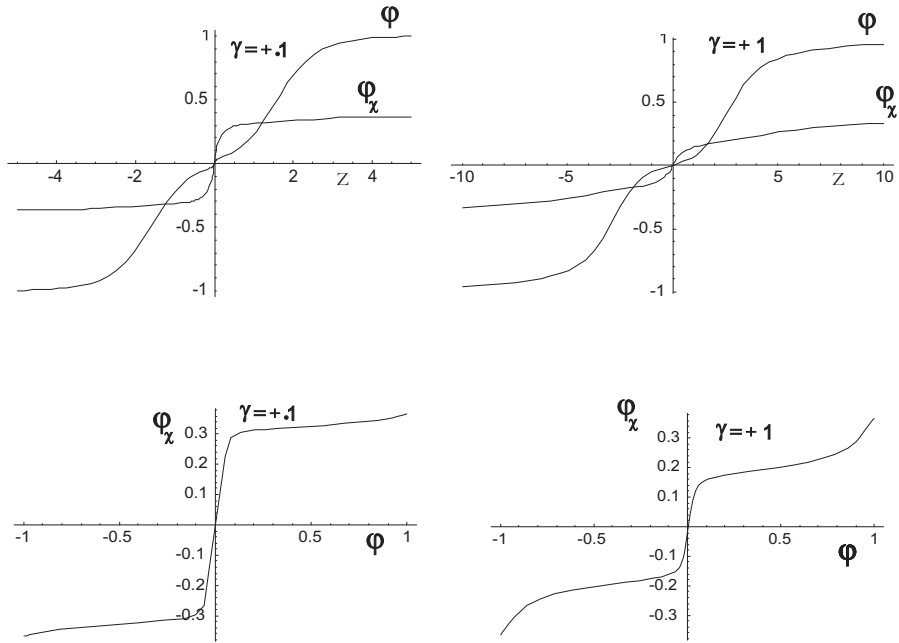} 
\end{center} 
\caption{Explicit domain wall solutions obtained setting 
$\alpha=3$. On the top part of the panel we display  the  
solutions for the fields while the associated parametric curves are drawn below. We have considered the case in which  
the glueball field is heavier (for $\gamma = 0.1$) on the left side of the panel. The case in which the glueball is lighter is presented  
on the right side of the panel (for $\gamma =1$).}\label{figureDOMAIN} 
\end{figure}

We also have 
\begin{eqnarray} 
 \partial_{{S}}{{W}}=\frac{2N^2}{3}\,\log_{(0)}  
\left(\frac{S}{-e\frac{\chi}{N} \log_{(0)} \chi^N}\right)\ ,\qquad  \frac{\partial_{\chi}{W}}{S}=\frac{2N}{3} \, \frac{\log_{(0)} \chi^N + N}{-\frac{\chi}{N}\log_{(0)}  \chi^N}\,. 
\end{eqnarray} 
 It is now possible to deduce a number of constraints. First of all
the invertibility of the \KA~metric  requires $h$ not to be factorizable, {\it  i.e.} it cannot be of the form $h(\chi ,\bar{\chi}) = {\cal {F}}(\chi)\, \overline{\cal {F}}(\bar{\chi})$ with arbitrary ${\cal F}$. This condition is more easily understood when looking at the denominator in (\ref{DE}) which vanishes identically when $h$ is factorizable. We also require $h$ not to modify the vacuum structure dictated by the superpotential of the theory. In this section we will analyze in detail the $N=2$ case and hence the solution must interpolate between the minimum at $\varphi =1 $ and $\varphi_{\chi}=1/e$ and the one at $\varphi =-1 $ and $\varphi_{\chi}=-1/e$. It is the origin, in the field space, which causes problems. This is clear in the analysis \cite{Kovner:1997im,{Kovner:1997ca}} reviewed earlier in the paper and when analyzing the solution only in terms of $\chi$.  
To avoid that the solution stops at the origin of the field space one 
must use the field $\chi$ to eliminate the singular behavior of the 
\KA~at that point due to the field $S$.   
A \KA~which is not factorizable, removes the singular behavior at the origin of the moduli space and is consistent with the analysis of the previous sections is~\footnote{We have also explored other possibilities for the \KA~potential which are reported in the appendix.} 
\begin{equation} 
\label{kaehler1}
h(\chi, \bar{\chi}) =(1+\gamma\,e^2\,\chi\bar{\chi} ) (e^2\,\frac{\chi\bar{\chi}}{N^2}\log{\chi^N}\log\bar{\chi}^N)^{-\frac{1}{3}}\ . 
\end{equation} 
Before presenting the numerical solution it is instructive to show how the previously introduced \KA~resolves the problem at  
the origin of the moduli space. Consider the term   
\begin{eqnarray} 
\lim_{\varphi,\varphi_{\chi}\rightarrow 0}\frac{(\bar{\varphi}\varphi)^{\frac{2}{3}}}{h\, h_{\chi\bar{\chi}} - h_{\chi}\bar{h}_{\bar{\chi}}}\, h_{\chi\bar{\chi}} \propto \frac{1}{9\gamma}  \ ,   
\end{eqnarray} 
where in the last step we have taken the limit in which $\varphi$ and $\varphi_{\chi}$ approach the origin simultaneously and neglected logarithmic corrections in the fields.  
It is not surprising that this ratio is proportional to $1/\gamma$ since $\gamma$ is responsible for the fact that the \KA~is not factorizable.  
All the other matrix elements exhibit analogous behavior near the origin. 
More interestingly the dimensionless parameter $\gamma$ is linked to the mass of the field $\chi$ which,  
after having integrated out the field $S$, reads 
\begin{eqnarray} 
M_{\chi} = \frac{2}{27}\,{\frac{\alpha}{\gamma}} \ . 
\end{eqnarray}  
The mass of $S$ after having integrated out $\chi$ is instead 
\begin{eqnarray} 
M_{S} = \frac{2}{3}\,\frac{\alpha}{(\gamma+1)} \ . 
\end{eqnarray} 
With this choice for the \KA~(see Appendix B for another choice) it 
turns out that the origin becomes problematic when the glueball mass 
diverges compared to the mass of the gluinoball. The glueball field 
must be present to construct explicit domain wall solutions in pure 
SYM. This is consistent with the claim of several authors in the 
literature. We can also consider two regimes, one in which the 
glueball mass is larger than the gluinoball mass, roughly for $\gamma 
<1/8$, and the other in which the glueball is the lightest 
state~\footnote{The analysis in \cite{Feo:2004mr} shows that the 
mixing between the states is small.}. Although we have already argued 
in \cite{Feo:2004mr}  
that the glueball mass is larger than the gluinoball mass, it is interesting to show the domain wall solutions in both cases in figure \ref{figureDOMAIN} for the case of two colors.    
 
The complete domain wall solution due to the interplay of the two fields is more complicate than in the illustrative case considered in the previous section in which we eliminated $S$ via its equation of motion.  
{}For example we see that the glueball field reaches first the 
asymptotic values when it is more massive than the gluinoball, while this is not the case when the gluinoball is lighter. 
Our results reinforce the argument given in the literature \cite{Kogan:1997dt} that the domain wall  
solutions are very sensitive to the presence of heavy states. This is so since the integrating out  
procedure can strongly modify the potential landscape in which the 
domain wall solutions must be found.  
To better visualize the problem related to integrating out $\chi$ in the extended VY theory we restrict ourselves to the  
two color theory  while the results are general. Due to the form of the extended VY superpotential, integrating out $\chi$ leads to 
\begin{eqnarray} 
\chi = \pm \frac{1}{e} \ . 
\end{eqnarray}     
Thus $\chi$ is fixed in one of the two vacua, according to the sign of 
the field $S$. {}So for $S$ positive (negative), we have 
$\chi=+(-)1/e$. This means that after having integrated out $\chi$ the 
superpotential is just the VY theory. The \KA~ for $S$ is then the VY one 
and hence the walls interpolating between the desired vacua 
disappear. On the contrary, when integrating out $S$, due to the 
functional relation between $S$ and $\chi$, one is still able to 
construct the relevant wall solutions between the two correct vacua of 
the theory. 
 
Our analysis completes the proof of the existence of the domain walls 
within the extended VY theory while strengthening the widespread 
expectation of the presence of such objects in SYM. Moreover, it confirms that the VY theory must be extended  
to describe a number 
of relevant properties of SYM such as the domain wall solutions.

 \section{On the Generic BPS Domain Walls} 
 We have till now discussed BPS domain wall solutions which interpolates between two vacua connected by a straight line passing through the origin of the moduli space. We have also investigated non BPS domain wall solutions connecting two generic vacua. 
However we have not yet discussed the generic BPS solutions 
connecting any two vacua not aligned with the origin. These domain walls are also known to be problematic to obtain via the simple VY theory \cite{Kogan:1997dt}. Indeed the VY superpotential has branch cuts in between the vacuua leading to cusps in the potential of the theory. Hence a domain wall solution, in the present case, will have to pass through these cusps with inevitable shortcomings. This situation has been discussed in detail in \cite{Kogan:1997dt}. The problem here is of different nature than the one for the domain walls passing through the origin of the moduli, that was essentially due to the \KA~ of the VY theory. However, it is natural to ask if also this problem can be solved or alleviated within our framework. Although we have not yet found a satisfactory solution we present our current understanding which we hope can help shedding some light on the problem.

We start by noticing that the extended VY theory has another interesting property, i.e. the discontinuities in the $S$ plane have been removed due to the presence of the glueball field $\chi$. As we shall see below this is similar, in spirit, to the case of one flavor super QCD \cite{Smilga:1997cx,{Kovner:1997ca},{deCarlos:1999xk}} but with important differences. The discontinuities reappear though in the $\chi$ plane. In principle this situation is more appealing since one can imagine to find domain walls connecting any two vacua not aligned with the origin of the moduli without crossing the logarithmic branch. In particular one would like to find solutions interpolating between the vacua labelled by two distinct integers, i.e. $k$ and $k'$. To analyze this situation one can consider the condition (see equation (\ref{imaginary}) in appendix A)
\begin{equation}
\Im m\left[e^{i\beta}W\right]~=~\mbox{\rm constant}\ ,
\end{equation}
where the superpotential is evaluated on the solution of the equations of motion. 
%
{}For domain walls connecting two vacua which are on a straight line crossing the origin the previous constraint is easily satisfied and it can be reduced to the problem investigated in the previous sections.
Unfortunately we have not yet found the generic BPS solutions connecting generic vacua not aligned with the origin. One of the reasons why we think that such solutions may exist is that the superpotential is very similar to the Taylor, Veneziano and Yankielowicz (TVY) superpotential for one massless flavor super QCD \cite{Taylor:1982bp}. This is even more transparent if one defines 
\begin{eqnarray}
Q=-\frac{\chi}{N}\log\chi^N \ .
\end{eqnarray}
 and rewrites the superpotential as:
\begin{eqnarray}
W=\frac{2}{3}N^2\,S\left[\log\frac{S}{Q}-2\right] \ .
\end{eqnarray} 
This similarity, however, is only an apparent one. In fact there are many differences: i) Our original superpotential saturates correctly the anomalies of super Yang-Mills and not the ones of super QCD. 
ii) In the original variables the superpotential has well defined vacua while in massless super QCD the vacuum runs away. 

{}From the generic domain wall solutions point of view one might still be tempted to say that the problem has been reduced to finding domain wall solutions in the super QCD effective theory a l\'{a} TVY. The domain wall solutions were found for this theory in \cite{Kovner:1997ca,{Smilga:1997cx},{deCarlos:1999xk}} for a nonzero matter field mass. Unfortunately we cannot simply borrow these solutions. In our case, in the variable Q, we have zero mass. Besides we are using a different \KA~\footnote{We remind the reader that the \KA~ used in the present work does not spoil any of the anomalies of the underlying gauge theory and it allows for domain wall solutions in between vacua aligned with the origin of the moduli.} than the one used in literature for super QCD. We will further investigate these type of domain wall solutions elsewhere.

\section{Conclusions} 
 
We investigated the vacuum structure of $SU(N)$ ${\cal N}=1$ super 
Yang-Mills. The theory is expected to possess $N$ vacua with 
associated domain walls. However, when using just the 
Veneziano-Yankielowicz  effective theory the domain wall solutions 
among these vacua were not found. We showed that the newly extended version of the low energy effective  
Lagrangian supports the expected domain wall solutions. More specifically we find the domain wall solutions associated with two vacua which are 
aligned with the origin of the moduli space. For the two color theory our analysis exhaust the number of possible BPS domain wall solutions. For a generic theory with even $N$ we can construct $N/2$ BPS domain wall solutions. These are the ones for which the related vacua are on a straight 
line crossing the origin. It turns out that the tension of these domain walls, at large $N$, is quadratic in $N$, in agreement with general expectations for this kind of solitons.  

We constructed new non-BPS domain wall solutions, for a generic $SU(N)$ SYM theory associated with any two vacua. 
These have tension scaling as $N^2$ and in some cases they reduce to the BPS ones described above.  Finally
 we have discussed the generic BPS domain walls related to two vacua not aligned with the origin. We have also suggested an intriguing relation
 between the domain walls description of SYM via the extended VY theory and similar domain walls for super QCD within the TVY effective theory \cite{Taylor:1982bp}.
  
A number of relevant puzzles remain unsolved, such as the BPS domain wall solutions associated with any vacua not aligned with the origin of the moduli space. However our analysis shows for the first time how domain walls emerge within the effective theory supposed to describe the low energy dynamics of pure ${\cal N}=1$ SYM.  
Our results are also a further test of the extended VY theory. Among possible interesting applications one might consider the effects of matter fields.      
 
\vskip 1cm \centerline{\bf Acknowledgments} We thank M. 
Shifman and A.V. Smilga for useful discussions.  F.S. thanks A.V. Smilga for suggesting this interesting problem while visiting the Institute of Theoretical Physics in Minnesota.

\appendix 
 
\section{BPS Domain Walls for a Generic Supersymmetric Sigma Model} 
In the paper of Chibisov and Shifman \cite{Chibisov:1997rc} one can find a very detailed and exhaustive discussion of domain walls for supersymmetric theories. Here, to make the paper self-contained, we re-determine the domain wall solution for a generic supersymmetric sigma model, contemplating a general \KA~for the theory.  
Consider a generic $\mathcal N=1$ supersymmetric sigma model defined 
by the Lagrangian  
\eq\label{lagrangian N=1} 
\begin{aligned} 
\mathcal L&=\int\dd^4\theta\;\mathcal K[\Phi^i,\overline{\Phi}^{\bar \imath}]+ 
\left[\int\dd^2\theta\; W[\Phi^i]+{\rm c.c.}\right]\\ 
&=-g_{i\bar \jmath}\de_\mu\varphi^i\de^\mu\bar\varphi^{\bar 
\jmath}+F^ig_{i\bar \jmath}F^{\bar \jmath}+(F^i\de_iW+{\rm c.c.})+\text{fermions}\;, 
\end{aligned} 
\fine where as usual  
\eq  
\label{metric} 
g_{i\bar\jmath}=\frac{\de^2\mathcal 
K}{\de\varphi^i\de\bar\varphi^{\bar \jmath}} \ .\fine Here the generic 
superfield  reads in components $\Phi^i=\varphi^i+\sqrt 
2\theta\psi^i+(\theta\theta)F^i$ $(i=1,..,D$ and the bar means complex 
conjugation). 
 
Eliminating the auxiliary fields using their equations of motion,  
$F^i=-g^{i\bar\jmath}\de_{\bar\jmath} \overline W$  
up to fermions, we obtain for the bosonic Lagrangian the following expression 
\eq \mathcal L=-g_{i\bar 
\jmath}\de_\mu\varphi^i\de^\mu\bar\varphi^{\bar\jmath} 
-V(\varphi^i,\bar\varphi^{\bar\jmath}) \;, 
\fine 
where 
$V=\de_iWg^{i\bar\jmath}\de_{\bar\jmath}\overline W$. 
 
The theory may support domain walls: these are classical solutions of 
the equations of motion with 
finite tension $\mathcal E=\int^{+\infty}_{-\infty}\dd z\,\varepsilon$. The fields can be chosen 
to depend only on one coordinate, say  
the coordinate $z$. 
The energy density of the Lagrangian \eqref{lagrangian N=1} in this case is given by 
\eq 
\begin{aligned} 
\varepsilon&=g_{i\bar\jmath}\frac{\dd\varphi^i}{\dd 
z}\frac{\dd\bar\varphi^{\bar\jmath}}{\dd z}+\de_iWg^{i\bar\jmath} 
\de_{\bar\jmath}{\overline W}\\ 
&=g_{i\bar\jmath}\left[\left(\frac{\dd\varphi^i}{\dd z}-e^{-i\beta} g^{i\bar 
k}\de_{\bar k}{\overline W}\right)\left(\frac{\dd\bar\varphi^{\bar\jmath}}{\dd z} -e^{i\beta} 
g^{\bar\jmath\ell}\de_{\ell}W\right)\right]+2\frac{\dd}{\dd z}\Re \text e[e^{i\beta}W]\;, 
\end{aligned} 
\fine  
where in the last step $\Re\text e$ means the real part and we 
introduced an arbitrary phase $e^{i\beta}$. Finiteness of $\mathcal E$ requires the boundary condition 
$\varepsilon (z=\pm\infty)=0$. Since the kinetic and potential terms 
are both positive definite at spatial infinity, the previous condition 
requires the fields $\varphi^i$ to approach  a constant 
value $\varphi^i_0$ (the vacua of the theory) at infinity such that 
$V(\varphi^i_0,\bar\varphi^{\bar\imath}_0)=0$ and $\frac{\dd\varphi^i}{\dd 
z}(\pm\infty)=0$. 
 
For a given boundary condition (say 
$\varphi^i(+\infty)=\varphi^i_0$, $\varphi^i(-\infty)=\varphi^i_1$), 
the total energy is minimized if one finds a solution to 
the following first order differential equation: 
\eq\label{BPS solution}  
\frac{\dd\varphi^i}{\dd z}=e^{-i\beta} 
g^{i\bar\jmath}\de_{\bar\jmath}{\overline W} \ .\fine 
One can show that this solution is BPS and the 
tension is simply:
\eq \label{tension} 
\mathcal E 
= 2\left\{\Re\text e[e^{i\beta} W(\varphi^i_0)]-\Re\text e[e^{i\beta}W(\varphi^i_1)]\right\} \ .  
\fine  
Such a 
solution can be also shown to preserve 1/2 of the original supersymmetry. 
This can be seen considering the general supersymmetry variation (for 
vanishing fermionic fields)  
\eq  
\delta_\epsilon\psi^i=\sqrt 
2\left[i(\sigma^3\bar\epsilon)\frac{\dd\varphi^i}{\dd 
z}+F^i\epsilon\right] \;, 
\fine  
while the other variations vanish 
identically. If we choose a Weyl spinor $\epsilon_\alpha$ such 
that $\epsilon_\alpha= 
ie^{-i\beta}\sigma^3_{\alpha\dot\gamma}\bar\epsilon^{\dot\gamma}$, the variation of the fermionic fields is exactly zero on the BPS solutions \eqref{BPS solution}, as it can be easily shown by using the  equation of motion of $F^i$. 
 Another well known useful relations can be found by multiplying the linear differential equation by $\partial_{i}W$, after noticing that $\beta$ is a constant one immediately derives:
\begin{eqnarray}
\frac{d\left[e^{i\beta}W\right]}{dz}=V \ .
\end{eqnarray}
Since the potential of the theory $V$ is real one deduces
\begin{eqnarray}
\Im m \left[e^{i\beta}W\right] = ~{\rm constant} \ ,\label{imaginary}
\end{eqnarray} 
which is valid when evaluated on the equations of motion.
There may exist also non-BPS domain wall solutions.  
These are classical solutions of the standard equation of motion  
\eq\label{non BPS} 
\frac{\dd}{\dd z}\left[g_{i\bar\jmath}\frac{\dd\bar\varphi^{\bar\jmath}}{\dd 
z}\right]=\frac{\de V}{\de\varphi^i} \,. \fine  
This equation admits the first integral 
of motion (conserved charge) \eq Q=g_{i\bar\jmath}\frac{\dd\varphi^{i}}{\dd z} 
\frac{\dd\bar\varphi^{\bar\jmath}}{\dd 
z}-V= {\rm constant} \ . \fine  
Equations \eqref{non BPS} can be interpreted 
as the motion of a point particle constrained on a $D$-dimensional K\"ahler manifold 
with  metric $g_{i\bar\jmath}$ and subject to the potential 
$-V$. The conserved charge $Q$ is the energy of such a particle. The 
boundary conditions  
for the domain wall solutions, namely that  
for $z\rightarrow\infty$ both the kinetic and the potential energy must vanish, imply $Q=0$.

\section{A different K\"{a}hler} 
\begin{figure}[htbp] 
\begin{center} 
\includegraphics[scale=.6]{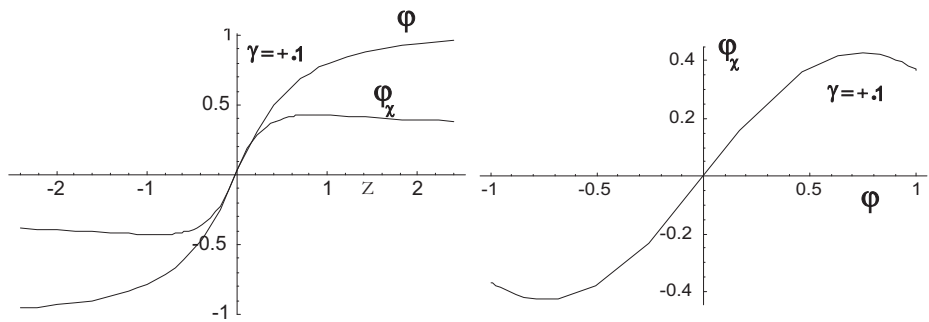} 
\end{center} 
\caption{Domain wall solutions obtained setting $\alpha=3$ and $\gamma= 0.1$.}\label{figureDOMAIN2} 
\end{figure}

Another simple \KA~which is not factorizable and solves the singular 
behavior at the origin of the moduli space is 
\begin{equation} 
h(\chi, \bar{\chi}) =(1+\gamma\,e^2\,\chi\bar{\chi} ) (e^2\,\chi\bar{\chi})^{-\frac{1}{3}}\ . 
\end{equation} 
We consider the term:   
\begin{eqnarray} 
\lim_{\varphi,\varphi_{\chi}\rightarrow 0}\frac{(\bar{\varphi}\varphi)^{\frac{2}{3}}}{h\, h_{\chi\bar{\chi}} - h_{\chi}\bar{h}_{\bar{\chi}}}\, h_{\chi\bar{\chi}} = \frac{1}{9\gamma} \, \frac{(\bar{\varphi}\varphi)^{\frac{2}{3}}}{(e^2\bar{\varphi}_{\chi}\varphi_{\chi})^{\frac{2}{3}}} \propto \frac{1}{\gamma}  \ , 
\label{hhh}\end{eqnarray} 
where in the last step we have taken the limit in which $\varphi$ and 
$\varphi_{\chi}$ approach the origin simultaneously. This ratio is 
also proportional to $1/\gamma$ since $\gamma$ is again responsible 
for the fact that our \KA~is not factorizable.  The dimensionless 
parameter $\gamma$ is linked to the mass of the fields $\chi$ and $S$ 
in the following way 
\begin{eqnarray} 
M_{\chi} = \frac{2\alpha}{3(1+4\gamma)} \ , \qquad M_{S} = \frac{2\alpha}{3(1+\gamma)} \ . 
\end{eqnarray}  
{}Here the singular behavior in (\ref{hhh}) happens when the glueball and the gluinoballs have the same (finite) mass and hence it makes this \KA~somewhat less physical than the one used in the main test. Actually, as it happens for the \KA~in eq.~\re{kaehler1}, the singularity should more naturally occur when the glueball field decouples.
However we can still solve for the domain walls which we show in figure \ref{figureDOMAIN2} for the case $N=2$.   
In this case the two masses are roughly comparable, with the glueball 
mass slightly lighter than the gluinoball mass.

\end{document}